# Water Retention of Rigid Soils from a Two-Factor Model for Clay

V.Y. Chertkov*

*Division of Environmental, Water, and Agricultural Engineering, Faculty of Civil and Environmental Engineering, Technion, Haifa 32000, Israel*

**Abstract**: Water retention is one of the key soil characteristics. Available models of soil water retention relate to the curve-fitting type. The objective of this work is to suggest a physical model of water retention (drying branch) for soils with a rigid matrix. "Physical" means the prediction based on the a priori measured or estimated soil parameters with a clear physical meaning. We rely on the two-factor model of clay that takes into account the factors of capillarity and shrinkage. The key points of the model to be proposed are some weak pseudo shrinkage that the rigid soils demonstrate according to their experimental water retention curves, and some specific properties of the rigid grain matrix. The three input parameters for prediction of soil water retention with the rigid grain matrix include inter-grain porosity, as well as maximum and minimum grain sizes. The comparison between measured and predicted sand water retention curves for four different sands is promising.
**Keywords**: Water retention; Rigid soils; Pseudo shrinkage; Capillarity and shrinkage factors; Non-fitting model

## 1. INTRODUCTION

The possibility of predicting an observed soil water retention curve from a number of physical soil parameters that are measured or estimated independently of the soil water retention, is so far lacking, even in the approximation of a *rigid* soil matrix. In this work we are only interested in this last case. Available models of the water retention in rigid soils (e.g., [1-10]) are eventually reduced to curve-fitting to relevant experimental soil water retention data. At least a part of the parameters used in the fitting in each of the models has no clear physical meaning and can only be found by fitting. Although the models can be practically useful for applications in soil technology and water management, their possibilities from the viewpoint of advancement in physical understanding and knowledge of the links between soil structure and soil water retention as a function of the structure, are in the best case, limited.

The objective of this work is to partially "dilute" the curve-fitting domination and to suggest some physical alternative as applied to the consideration of water retention (drying branch) in *rigid* soils. The attempt to be proposed relies on the concepts and results of a recent work devoted to pure clay water retention [11]. Such *paradoxical* relations between the model for *clay* and water retention of *rigid* soils at first glance seem strange. However, as will be shown, they flow out of some *pseudo shrinkage* property of rigid soils. The physical meaning of "pseudo shrinkage" in the case of rigid soils will be explained below (Section **4.1**). Here, it is just worth noting that this "pseudo shrinkage" has no relation to the true shrinkage of clay or clay soils.

We consider soils that are rigid as a whole, but shrinkage of the small clay clusters in the inter-grain pore space, i.e., micro-shrinkage can take place. Such soils can be considered as a system of silt and sand grains (see [12,13] for a brief discussion of the aggregated soil transition to such a system with the clay content striving to zero). Meaning such soils, for brevity we use the term "sand". Although we only consider rigid soils, possible applications of the to-be-obtained results to shrinking soils are briefly indicated in Section **6**.

Some relations of the model [11] that are necessary in the following are presented in the beginning of the exposition. We also emphasize some points from [11] that should be modified to reach the objectives of this work. Notation is summarized at the end of the paper.

## 2. SOME NECESSARY RELATIONS OF THE CLAY TWO-FACTOR MODEL

The soil suction $h$ can be presented as a product of two factors

$$h = HQ \quad . \tag{1}$$

---

*Address correspondence to this author at the Division of Environmental, Water, and Agricultural Engineering, Faculty of Civil and Environmental Engineering, Technion, Haifa 32000, Israel; E-mail: agvictor@tx.technion.ac.il

The *H* factor originates from the adsorption-capillary phenomena. The *Q* factor originates from the shrinkage-swelling of the soil matrix. In the particular case of a pure clay *h* only depends (through *H*) on one characteristic size of water-containing pores or pore tubes ("water-containing" implies different degrees of water filling). This single characteristic size can be generalized to the case of a rigid matrix accounting for its specifics (see Section **4.6**).

In the case of a clay *Q* is found as

$$Q=(1-v(\zeta))^2/(1-v_z)^2, \qquad 0<\zeta<1 \qquad (2)$$

where *v* is the relative clay volume (see the Notation); $\zeta$ is the relative water content of clay (see the Notation); $v_z \equiv v(\zeta_z)$ is the *v* value at the shrinkage limit of the clay, $\zeta=\zeta_z$. Note that $v(\zeta)$ is known from the physical model of the clay shrinkage curve [14,15]. In Eq.(**2**) $\zeta=1$ is considered to be the initial point of clay shrinkage. With that $v(\zeta)$ is in the range $v_z \leq v < 1$.

In the range $\zeta_* < \zeta < 1$ *H* for clay is only connected with capillarity as

$$H=4\Gamma\cos\alpha_c/R(\zeta), \qquad \zeta_*<\zeta<1 \quad . \qquad (3)$$

Here $\zeta_* \cong 0.1\zeta_z$ (that is higher than residual water content) is an upper estimate of the lower $\zeta$ boundary ($\zeta_*$ can be lower than this estimate, to $\zeta_* \cong 0.05\zeta_z$, see e.g., Table 1 of [15]); $\Gamma$ is the surface tension of water; $\alpha_c$ is a contact angle; and $R(\zeta)$ is a characteristic internal size of pore tubes of the clay matrix at a cross-section.

The $R(\zeta)$ size is written as (Fig.(**1**))

$$R(\zeta) = \begin{cases} \rho'_m(\zeta), & \zeta_n < \zeta \leq 1 \\ \rho'_c(\zeta), & \zeta_* \cong 0.1\zeta_z < \zeta \leq \zeta_n \end{cases}, \qquad (4)$$

where $\rho'_m(\zeta)$ (Fig.(**1**), curve 2) is the maximum internal size of pore tube cross-sections in the $\zeta_n<\zeta<1$ range ($\zeta=\zeta_n$ is the clay air-entry point); $\rho'_c(\zeta)$ (Fig.(**1**), curve 4) is the maximum internal size of the water-containing pore tubes in the $\zeta_* < \zeta < \zeta_n$ range. The *H* presentation in Eq.(**3**) reflects the physical peculiarity of a clay matrix structure. At least in the area of normal shrinkage, $\zeta_n<\zeta<1$ there is only one characteristic size - the maximum internal size of pore-tube cross-sections $\rho'_m(\zeta)$ (Fig.(**1**), curve 2), that coincides with the maximum internal size $\rho'_f(\zeta)$ of the water-filled pore tubes in this area.

In the area $\zeta_n<\zeta<1$ $R(\zeta) \cong \rho'_m(\zeta)$ (Eq.(**4**); Fig.(**1**), curve 2) is expressed through $v(\zeta)$, $v_z$, $v_s$ (the relative volume of clay solids; see the Notation), $r_{mM}$ (the maximum external size of clay pores at $\zeta=1$); and characteristic constants of the clay microstructure, $\alpha$ and *A* [14]. In the area $\zeta_*<\zeta<\zeta_n$ $R(\zeta) \cong \rho'_c(\zeta)$ (Eq.(**4**); Fig.(**1**), curve 4) is found to be a solution $\rho'_c(\zeta)$ of the water balance equation (at a clay cross-section) as

$$F(\zeta)=\varphi(\rho'_f)+\int_{\rho'_f}^{\rho'_c} g(\rho')\frac{d\varphi}{d\rho'}d\rho', \qquad \zeta_*<\zeta<\zeta_n \qquad (5)$$

where *F* is the saturation degree at a relative water content $\zeta$; $\varphi(\rho')$ is the pore tube-size distribution; $\rho'_f(\zeta)$ is the maximum internal size of the water-filled pore-tube cross-sections in the area $\zeta_*<\zeta<\zeta_n$ (Fig.(**1**), curve 3); and $g(\rho')$ is the degree of filling of the pore tubes of internal $\rho'$ size with water ($0<g<1$). The first and second terms in the right part of Eq.(**5**) give the contributions of the water-filled and water-containing pores, respectively. For clay $F(\zeta)$ in Eq.(**5**) is found to be [14]

$$F(\zeta)=[(1-v_s)/(v(\zeta)-v_s)]\zeta, \qquad 0<\zeta<1 \quad . \qquad (6)$$

The expression for $\varphi(\rho')$, and the details for solving Eq.(**5**) to find $\rho'_c(\zeta)$ should be modified compared to [11] (see Sections **3.2** and **3.3**) accounting for the specifics of the rigid grain matrix (Section **4.6**).

The $\rho'_c(\zeta)$ solution of Eqs.(**5**) and (**6**) determines $R(\zeta)$ (Eq.(**4**)) and *H* (Eq.(**3**)). The final expression for the clay suction, $h(w)$ (Eqs.(**1**)-(**3**)) is given by

$$h(w)=[4\Gamma\cos\alpha_c/R(w/w_M)] \, [1-v(w/w_M)]^2/(1-v_z)^2, \qquad w_* < w < w_M \qquad (\zeta_* < \zeta = w/w_M < 1) \,. \tag{7}$$

The input physical parameters of the model are $v_s$, $v_z$, $r_{mM}$, and the density of clay solids, $\rho_s$. However, $r_{mM}$ is connected with the maximum size of clay particles in the oven-dried state, $r_{mz}$ as $r_{mM}=r_{mz}v_z^{-1/3}$ [14]. If we take $r_{mz}\cong 2\mu m$ (according to the generally accepted definition of the maximum size of clay particles in the oven-dried state) $r_{mM}$ is estimated to be $r_{mM}\cong 2v_z^{-1/3}$ (μm). This result is used in Section **4.6**. Some approximations of the two-factor model should be specified.

## 3 THE NECESSARY MODIFICATIONS OF THE CLAY TWO-FACTOR MODEL

### 3.1 Accounting for the More Accurate Maximum Swelling Point of Clay

The above two-factor model neglects the difference between the maximum swelling point of clay, $w_h$ and the clay liquid limit, $w_M$. This approximation ($w_h\cong w_M$, i.e., $\zeta_h\cong\zeta_M=1$) influences the $Q$ factor (Eq.(**2**)) and changes the clay suction $h$ in the $\zeta_n<\zeta<\zeta_h$ range compared to a real case when $w_h<w_M$, $\zeta_h<1$ (see $\zeta_h$ in Fig.(**1**)). The interrelation between $w_h$ and $w_M$ for clay as

$$w_h\cong 0.5w_M \qquad (\zeta_h=w_h/w_M\cong 0.5) \tag{8}$$

was obtained recently [16,17]. Following the derivation of Eq.(**2**) [11], in the case where $w_h\cong 0.5w_M<w_M$ (i.e., at $\zeta_h\cong 0.5$), one should replace $1=v_M=v(\zeta_M=1)$ in the numerator and denominator of Eqs.(**2**) and (**7**) with $v_h=v(\zeta_h\cong 0.5)$ and the $0<\zeta<1$ range with the $0<\zeta<\zeta_h\cong 0.5$ range (see $\zeta_h$ in Fig.(**1**)). Using the clay shrinkage curve, $v(\zeta)$ [14,15] gives $v_h=v(\zeta_h\cong 0.5)=0.5(v_s+1)$. Thus, in the modified model Eq.(**2**) is replaced with

$$Q=(v_h-v(\zeta))^2/(v_h-v_z)^2, \qquad 0<\zeta<\zeta_h\cong 0.5 \,. \tag{9}$$

The final expression for $h(w)$ (Eq.(**7**)) is modified as

$$h(w)=[4\Gamma\cos\alpha_c/R(w/w_M)] \, [v_h-v(w/w_M)]^2/(v_h-v_z)^2, \qquad w_* < w \leq 0.5w_M \qquad (\zeta_* < \zeta < \zeta_h) \,. \tag{10}$$

### 3.2. More Convenient Presentation of Pore Size Distribution

The form of the presentation of a pore-size distribution plays an important role. Chertkov [14,15] used the presentation that is convenient for considering clay shrinkage. The convenience consists in the use of *external* pore ($r$) and pore tube ($\rho$) sizes (i.e., the sizes that include a half-thickness of clay particles limiting the pores). In this case the volume of any pore, that is proportional to $r^3$, is proportional to the clay volume at shrinkage. However, such a presentation does not include, in an explicit form, the clay porosity that is connected with *internal* pore sizes ($r'$ and $\rho'$) which determine the clay water retention. The generalization, giving a more convenient presentation of pore-size distribution, using internal pore sizes, and explicitly including porosity as a distribution parameter, was suggested recently [18]. In addition, this presentation in a natural way is generalized to a two- or multi-mode porosity case that can be topical for clay and soil. The modified presentation of the pore-tube size distribution $\varphi(\rho')$ will be indicated as applied to rigid soils in an explicit form in Section **4.6**.

### 3.3 Simplifying the Solution of the Water Balance Equation

Solving Eq.(**5**) in the clay two-factor model was based on some assumptions about $g(\rho')$ and pore shape. The solution can be simplified and specified. In the following consideration of a rigid grain matrix we rely on the above two-factor model for clay and, in particular, Eq.(**5**). In the case of a rigid soil, however, dependences $F(\zeta)$ and $\varphi(\rho')$ in this equation qualitatively and quantitatively differ. We consider the solving modification in Section **4.6** to be applied to a rigid soil.

## 4. WATER RETENTION OF A RIGID-GRAIN MATRIX

### 4.1. Specific Physical Features of the Rigid-Grain Matrix Compared with a Clay One

The presentation itself of the suction $h$ through the $Q$ and $H$ factors (Eq.(**1**)) is general [11]. However, in Sections **2** and **3.1** we essentially relied on the specific physical features of the shrink-swell network of clay particles. The specific features of the rigid grain matrix are also essential in the consideration of the $Q$ and $H$ factors for this case. Fig.(**2**) shows the general qualitative view of the $Q$ factor for any soil. Indeed, in some range, $0<\zeta \leq \zeta_z$ $Q=1$ (Fig.(**2**)) and in a point, $\zeta=\zeta_o$ (Fig.(**2**)) the suction $h(\zeta_o)=0$. Since the maximum pore size at $\zeta=\zeta_o$ usually remains a capillary one, we have $H(\zeta_o)\neq 0$ (Fig.(**2**)). Hence, $Q(\zeta_o)=0$ (Fig.(**2**)). Thus, $Q$ smoothly decreases to zero in the $\zeta_z<\zeta\leq\zeta_o$ range (Fig.(**2**)). In the case of a clay $\zeta=\zeta_z$ is the shrinkage limit, and $\zeta=\zeta_o\equiv\zeta_h$ is the maximum swelling point (for $\zeta_h$ see Section **3.1**). In the case of a sand we keep the same designations of the characteristic points $\zeta_z$ and $\zeta_o$ on the $Q(\zeta)$ curve (Fig.(**2**)), but their physical meaning, naturally, changes (see below).

Owing to the indicated qualitative similarity of the $Q(\zeta)$ curve for both a shrink-swell clay and rigid sand (Fig.(**2**)) and accounting for the $Q$ expression for a clay through the clay shrinkage curve (Eq.(**9**)), we can formally consider the $Q$ factor for a sand as originating from some "shrinkage" curve (like Eq.(**9**)) with a number of specific features (Fig.(**3**)). These features flow out of the simple generally known facts.

(i) The sand volume should not change with water-filling in its pores up to saturation (rigid matrix). This means that the range $0<\zeta\leq\zeta_z$ where $Q=const=1$ (Fig.(**2**)) and the relative volume $v=const=v_z$ (Fig.(**3**)), correspond to increasing the water content up to saturation at $\zeta=\zeta_z$. This condition (of water saturation) at $\zeta=\zeta_z$ (the *first specific feature*) can be written using the saturation degree, $F(\zeta)$ (Eq.(**6**); note that this expression for $F(\zeta)$ from [14] is suitable for any soil because the specific geometry of clay particles was not used in its derivation) as

$$F_z \equiv F(\zeta_z)=[(1-v_s)/(v_z-v_s)]\zeta_z=1 \tag{11}$$

(note that for a clay Eq.(**11**) is not true). The $v_s$ and $v_z$ parameters for sand formally have the same meaning as for clay. They are discussed in Section **4.2**. Thus, the physical meaning of $\zeta=\zeta_z$ for a sand matrix (Figs.(**2**) and (**3**)) is the water saturation point. In addition, regarding the rigid matrix as a boundary case of clay one can consider $\zeta=\zeta_z$ for sand to be the coincidence of two points, the shrinkage limit, $\zeta_z$ and the air-entry point, $\zeta_n$ (cf. the separate $\zeta_z$ and $\zeta_n$ positions for clay in Fig.(**1**); note that for clay $F(\zeta_n)=1$, but $F(\zeta_z)<1$, see Eq.(**6**) and [14]).

(ii) In any case, irrespective of the physical nature of the sand "shrinkage" in Fig.(**3**) (see below) the change of the relative volume, $1-v_z$ should obviously be very small. Thus, the *second specific feature* of sand (unlike clay) is (Fig.(**3**))

$$1-v_z \ll 1 \ . \tag{12}$$

(iii) Many data (e.g., Hillel [19], Fig.6.9, p.157) evidence that (unlike in clay or the clay soil case) the water retention of soils with a rigid matrix is characterized by a very steep suction decrease (sharp bend) down to zero in the small vicinity of water saturation. It follows that $\zeta_z$ and $\zeta_o$ (Fig.(**2**)) are very close ($\zeta_o-\zeta_z\ll 1$). Since in the case of sand $\zeta_z$ corresponds to the water saturation state (the *first specific feature*), the $\zeta_o$ water content (note, $\zeta_o>\zeta_z$) should have the maximum possible value, $\zeta_o=1$ (unlike clay for which $\zeta_o=\zeta_h<1$). Thus, the *third specific feature* of sand is (Figs.(**2**) and (**3**))

$$1-\zeta_z \ll 1 \ . \tag{13}$$

The physical nature and meaning of the small water content variation $(1-\zeta_z)$ and small sand volume variation $(1-v_z)$ in the area of the sand "shrinkage", $\zeta_z<\zeta\leq 1$ (Fig.(**2**) at $\zeta_o=1$ and Fig.(**3**)) flow out of the following considerations. Because $\zeta=\zeta_z$ is the water saturation point, the water addition, $1-\zeta_z$ (and corresponding sand volume addition, $1-v_z$) is that to the already saturated sand. That is, this water addition (unlike true shrinkage-swelling case) can only be in the form of some water film between the sand grains along their contact surfaces. The appearance of such an additional water film means some transition of the saturated sand to the state that should differ from the saturated one in the minimum possible degree because we consider the rigid matrix. This means that the $(1-\zeta_z)$ water addition should correspond to the film of a minimally possible characteristic thickness. The latter is discussed in Sections **4.4** and **4.5**.

## 4.2. The "Shrinkage Curve" of the Rigid-Grain Matrix

To explicitly present the sand "shrinkage" curve (illustrated in Fig.(**3**)), one should first express the $v_s$, $v_z$, and $\zeta_z$ parameters of the "shrinking" sand through values that are more convenient and relevant to describe the sand. As in the consideration of clay [14] we introduce the minimum volume of "shrinking" sand (in the area of the rigid matrix), $V_z$; the maximum volume, $V_M$; the volume of solid phase, $V_s$; and the volume of pores, $V_p$ in the area of the rigid matrix, $0<\zeta\leq\zeta_z$ (Fig.(**3**)). Additionally we introduce the total increment, $\Delta V$ of "shrinking" sand volume between $V_M$ and $V_z$ as

$$V_M = V_z + \Delta V \ , \tag{14}$$

and the total increment, $\Delta V_w$ of water volume associated with sand volume "shrinkage" between $V_M$ and $V_z$. Then, the relative additional sand volume, $\Delta v$ (per unit volume of dry sand or, in fact, the sand in the area of rigid matrix, $0<\zeta\leq\zeta_z$), and the relative additional water volume, $\Delta v_w$ (per unit volume of dry sand) are defined as

$$\Delta v \equiv \Delta V/V_z \ , \qquad \Delta v_w \equiv \Delta V_w/V_z \ . \tag{15}$$

Similar to the definition of the relative oven-dried volume, $v_z$ and relative solids volume, $v_s$ for clay [14,15] and accounting for Eqs.(**14**) and (**15**), for the "shrinking" sand (see Fig.(**3**)) one can write

$$v_z \equiv V_z/V_M = V_z/(V_z+\Delta V) = 1/(1+\Delta v) \tag{16}$$

and

$$v_s/v_z \equiv V_s/V_z = (V_z - V_p)/V_z = (1-p) \tag{17}$$

where $p$ is the (constant) sand porosity in the area $0<\zeta\leq\zeta_z$ (Fig.(**3**)). According to Eqs.(**16**) and (**17**)

$$v_s = (1-p)/(1+\Delta v) \ . \tag{18}$$

To find $\zeta_z$ one can use the *first specific feature* of the "shrinking" sand (Eq.(**11**)) with $v_z$ and $v_s$ from Eqs.(**16**) and (**18**), respectively. Then,

$$\zeta_z = p/(p+\Delta v) \ . \tag{19}$$

On the other hand, since $\zeta_z$ is the saturation point (see Section **4.1**) we directly have

$$\zeta_z = V_p/(V_p+\Delta V_w) = (V_p/V_z)/(V_p/V_z + \Delta V_w/V_z) = p/(p+\Delta v_w) \ , \tag{20}$$

and from Eqs.(**19**) and (**20**)

$$\Delta v = \Delta v_w \ . \tag{21}$$

Using Eqs.(**12**) and (**13**) one can easily show that the above specific features of the "shrinking" sand [$(1-v_z)\ll 1$ and $(1-\zeta_z)\ll 1$] can be presented in the following form

$$\Delta v_w \ll 1 \quad (\text{or } \Delta v \ll 1) \qquad \text{and} \tag{22a}$$

$$\Delta v_w/p \ll 1 \quad (\text{or } \Delta v/p \ll 1) \ . \tag{22b}$$

For this reason one can rewrite Eqs.(**16**), (**18**), and (**20**) in the linear approximation with respect to small value $\Delta v_w (=\Delta v)$ as

$$v_z \cong 1 - \Delta v_w, \qquad v_s \cong (1-p)(1-\Delta v_w), \qquad \zeta_z \cong (1-\Delta v_w/p) \ . \tag{23}$$

Note, that $\Delta v = 1 - v_z$ (see Fig.(**3**)) only in the linear approximation (cf. Eq.(**16**)), and $\Delta v_w \neq 1 - \zeta_z$ (see Fig.(**3**)) even in the linear approximation. Eq.(**23**) gives the simple expressions of $v_z$, $v_s$, and $\zeta_z$ through the more convenient parameters of "shrinking" sand, the usual sand porosity, $p$ and the small additional relative volume of the water films at sand grain contacts, $\Delta v_w$ (or the corresponding small additional relative sand volume, $\Delta v$). The $\Delta v_w$ is estimated below (Sections **4.4** and **4.5**).

Now, accounting for the physical conditions from Eqs.**12** and **13** one can write the small difference $(v(\zeta)-v_z)$ for the "shrinkage" curve $v(\zeta)$ of the sand in the very small range, $\zeta_z<\zeta\leq 1$ (Fig.(**3**)) as an expansion in powers of the small difference $(\zeta-\zeta_z)$ and be limited by the second power as (note that $v-v_z$ has the minimum at $\zeta=\zeta_z$)

$$v(\zeta) - v_z \cong k(\zeta-\zeta_z)^2 \ , \qquad \zeta_z < \zeta \leq 1 \ . \tag{24}$$

The condition, $v(1)=1$ (Fig.(3)) and $v_z$, $\zeta_z$ from Eq.(23) allow one to estimate the major term of the expansion of the $k$ coefficient in powers of the small $\Delta v_w$ value as

$$k \cong p^2/\Delta v_w \; . \tag{25}$$

Thus, the "shrinkage" curve of the sand is presented as (cf. Fig.(3))

$$v(\zeta) = \begin{cases} v_z, & 0 \leq \zeta \leq \zeta_z \\ v_z + (p^2/\Delta v_w)(\zeta - \zeta_z)^2, & \zeta_z < \zeta \leq 1 \end{cases}, \tag{26}$$

with $v_z$ and $\zeta_z$ expressed through $p$ and $\Delta v_w$ from Eq.(23).

### 4.3. The $Q$ Factor of the Rigid-Grain Matrix

According to Sections **4.1** and **4.2** the $Q$ factor for sand has the same form as for clay (Eq.(9)), but with the following essential modifications. In the case of sand $\zeta_o$ (Fig.(2)) is equal to 1 instead of $\zeta_h$ for clay. Correspondingly, for sand $v(\zeta_o)=v(1)=1$ instead of $v(\zeta_h)=v_h$ for clay. The "shrinkage" curve for sand, $v(\zeta)$ is given by Eq.(26) instead of $v(\zeta)$ for clay from [14,15]. Finally, $v_z$ for sand is given by Eq.(23). As a result the $Q$ factor for sand is given by Eq.(2) with $v(\zeta)$ from Eq.(26) and $v_z$ from Eq.(23). In the explicit form $Q$ is presented as

$Q(\zeta)=1$, $\qquad 0 \leq \zeta \leq 1-\Delta v_w/p$,

$Q(\zeta)=[1-(p/\Delta v_w)^2(\zeta-1+\Delta v_w/p)^2]^2$, $\qquad 1-\Delta v_w/p < \zeta \leq 1$. $\qquad (27)$

Equation (27) cannot be simplified in the area $1-\Delta v_w/p < \zeta \leq 1$ using $\Delta v_w/p \ll 1$ (Eq.(22)) because the variation range of $\zeta$, $\Delta\zeta = \Delta v_w/p$ is very small. The smaller the $\Delta v_w/p$ ratio is, the steeper is the decrease of $Q(\zeta)$ and $h(\zeta)(=HQ)$ for the sand at $\zeta > \zeta_z$ in Fig.(2).

To obtain the $Q$ factor for sand as a function of its gravimetric water content, $W$ one should replace $\zeta$ in Eq.(27) with the ratio, $\zeta = W/W_M$. The maximum value $W_M$ (see Fig.(2) with replacement $\zeta \to W$ and $\zeta_o \to W_M$) corresponds to $\zeta_o=1$ (Fig.(2)), i.e., to the state of the sand with additional water film of the minimum characteristic thickness along grain contacts. $W_M$ is estimated using the same expression as for the clay liquid limit [14], $W_M=[(1-v_s)/v_s](\rho_w/\rho_s)$ ($\rho_w$ - water density; $\rho_s$ - sand solid density). Using the specific expression for $v_s$ in the case of sand (Eq.(23)) and in the linear approximation with respect to $\Delta v_w$ one has

$$W_M=(\rho_w/\rho_s)(1+\Delta v_w/p)/[(1-p)/p] \; . \tag{28}$$

The gravimetric water content of the saturated state of the sand, $W_z$ (see Fig.(2) with replacement $\zeta \to W$ and $\zeta_z \to W_z$) corresponds to $\zeta=\zeta_z$ (Fig.(2)). In the lowest (square) approximation with respect to $\Delta v_w$ $W_z$ is written as

$$W_z = \zeta_z \, W_M = (\rho_w/\rho_s)[1-(\Delta v_w/p)^2]/[(1-p)/p] \; . \tag{29}$$

From Eqs.(28) and (29) one can estimate the $\Delta v_w/p$ ratio for the sand through $W_z$ and $W_M$ as

$$\Delta v_w/p \cong \Delta W/W_M \cong \Delta W/W_z \tag{30}$$

where $\Delta W = W_M - W_z$. The definitions of $\Delta v_w$, $p$, and $W$ also lead to Eq.(30).

For sand grains the volumetric water content, $\theta$ is more convenient. To pass from $\zeta$ to $\theta$ in Eq.(27) one can use the general relation between $\theta$ and $W$ as

$$\theta = (\rho_s/\rho_w)W[1-P(W)] \tag{31}$$

where by definition of $\zeta$ ($W_M$ from Eq.(28))

$$W = \zeta W_M, \tag{32}$$

and by definition of soil porosity, $P(W)$ the difference $[1-P(W)]$ is the solid volume to soil volume ratio. For clay (see [14]) or "shrinking" sand

$$1-P(W)=1-P(\zeta)=v_s/v(\zeta) , \qquad 0\leq\zeta\leq 1 . \tag{33}$$

Then, the relation,

$$\theta=(\rho_s/\rho_w) W_M [v_s/v(\zeta)]\zeta , \qquad 0\leq\zeta\leq 1 \tag{34}$$

permits one to recalculate the $\zeta$ values to $\theta$ ones, accounting for $v(\zeta)$ from Eq.(26), $v_z$, $v_s$, and $\zeta_z$ from Eq.(23), and $W_M$ from Eq.(28). Thus, one can find $\theta_z$ (at $\zeta=\zeta_z$, $v(\zeta)=v_z$) and $\theta_M$ (at $\zeta=1$, $v(\zeta)=1$) as well as the lower boundary of the model applicability, $\theta_*$ (see Fig.(2)) (at $\zeta=\zeta_*\sim 0.1\zeta_z$, $v=v(\zeta_*)$) that was noted in the lines before and after Eq.(3).

### 4.4. The Relative Additional Water Volume ($\Delta v_w$) of the Rigid-Grain Matrix

The objective of this section is to present the $\Delta v_w$ value as a simple function of a characteristic sand grain size ($X_m$) and characteristic minimum thickness ($l$) of the additional water film between the grains. The sand volume of cubic shape with side size, $L$ (Fig.(4a)) is considered to consist of layers of rounded grains. The layer thickness is taken to be equal to the mean grain size, $d=X_m/4$ (Fig.(4a)) where $X_m$ is the maximum grain size (see [18]). The grains in the cubic volume can be attributed to the three different layer systems that are normal to axes $x$, $y$, $z$, respectively (Fig.(4a)). The mean number of the layers in each the system is equal to the mean number of boundaries between the layers and given by the ratio, $L/d=L/(X_m/4)$ (Fig.(4a)). The total number of such boundaries in the cubic sand volume in all the three grain layer systems is $3L/(X_m/4)$. Figure (4b) shows (in the magnified view) the part of the cross-section, normal to a layer system in the water saturated state. One can see grains entering two vertical layers (with one boundary between them - the dashed line) and four horizontal ones (with three boundaries between them - the dashed lines) as well as water-filled inter-grain pores. Figure (4c) shows the same situation, but in the state when the water films of the minimum characteristic thickness, $l$ are added along the boundaries of the grain layers, moving the latter apart. Within the limits of the cubic sand volume the volume of each added water film of thickness $l$ and surface area $L^2$ is $l L^2$. We take the natural assumption, $l/X_m<<1$ (at least $l/X_m<$ or $\sim 0.1$) that will be justified below. Then, the total number of the water films in the sand cube (Fig.(4a)) is $3L/(X_m/4)$ (see above the number of boundaries between grain layers), and the total additional volume of the water films is $lL^2\ 3L/(X_m/4)=12(l/X_m)L^3$. As a result the relative additional water volume, $\Delta v_w$ per unit volume of the sand (including both grains and inter-grain pores) is simply estimated as

$$\Delta v_w=12\ l/X_m . \tag{35}$$

Using terms proportional to $(l/X_m)^2$ and $(l/X_m)^3$ this estimate can be specified. However, accounting for the above assumption, $l/X_m<<1$ and numerical estimates in Section **4.5** we can neglect these possible corrections in Eq.(35).

### 4.5. The Characteristic Thickness of the Additional Water Film ($l$)

The thickness, $l$ of the water films appearing between sand grains, in addition to the saturated state, should not be affected by gravity and capillarity. Indeed, there are only three characteristic lengths for such a grain system in water: $X_m$, the thickness $l_o$ of a monomolecular water layer ($\sim 3$Å), and the so-called capillary constant, $[2\Gamma/(\rho_w g)]^{1/2}$ (e.g., Landau et al. [20]; $\Gamma$ is the surface tension of water; $\rho_w$ is the water density; $g$ is the specific gravity) that reflects the possible joint effects of gravity and capillarity. For this reason the characteristic thickness $l$ of the water film can be presented, from dimension considerations, as

$$l=l_o\ \lambda(l_o/X_m, l_o/[2\Gamma/(\rho_w g)]^{1/2}) \tag{36}$$

where $\lambda$ is a function of the ratios of $l_o$ to $X_m$ and $[2\Gamma/(\rho_w g)]^{1/2}$, respectively. Since $l_o<<X_m$ (usually $X_m\sim 2$mm) and $l_o/[2\Gamma/(\rho_w g)]^{1/2}<<1$ (for water at $20^oC$ $[2\Gamma/(\rho_w g)]^{1/2}\sim 3.9$mm), we can, with high accuracy, be limited by the first constant term ($\lambda_o$) of the expansion of $\lambda$ in the powers of $l_o/X_m$ and $l_o/[2\Gamma/(\rho_w g)]^{1/2}$. Therefore, the characteristic thickness, $l\sim l_o\lambda_o$ of the additional water film practically does not depend on the characteristic grain size, $X_m$ as well as gravity and capillarity. This result is used below in the estimation of the $l$ value.

In estimating $l$ we rely on the consequences of known experimental facts and the analysis of the applicability conditions of Eq.(**35**). These consequences are as follows.

(i) As shown above, the known experimental facts relating to the rigid grain matrix lead to the condition, $\Delta v_w \ll 1$ (Eq.(**22**)).

(ii) On the other hand, in the case of the shrink-swell clay matrix this condition is obviously violated because in this case (also according to known experimental facts) $\zeta_o - \zeta_z \sim 1$ [or $(1-\zeta_z) \sim 1$] and $(1-v_z) \sim 1$ (cf. Section **4.1** and Figs.(**2**) and (**3**)). The violation of the condition $\Delta v_w \ll 1$ means that for the shrink-swell clay matrix the similar value, $\Delta v_w$ [the ratio of the water volume, $\Delta V_w$ that the clay loses when shrinking from the maximum swelling point to the shrinkage limit, to the clay matrix volume $V_z$ at shrinkage limit; see Eq.(**15**)] meets the following condition: $\Delta v_w >$ or $\sim 1$.

(iii) Still another known experimental fact states that in terms of grain (particle) sizes the rigid grain matrix ($\Delta v_w \ll 1$) and swell-shrink clay one ($\Delta v_w >$ or $\sim 1$), are separated each from other by some grain size band between $\sim 20$ and $\sim 2\mu$m.

As follows from the derivation of Eq.(**35**), this equation gives the *single-valued* connection between the $\Delta v_w$ value and the $l/X_m$ ratio, and relates to *rounded* grains (see Fig.(**4**)) that meet the condition $l/X_m \ll 1$ (at least $l/X_m <$ or $\sim 0.1$). Equation (**35**) was derived implying the silt-sand grain matrix, i.e., at least $X_m > 20\mu$m. However, we *assume* below that $l$ is so small (note, that $l$ does not depend on the $X_m$ size; see the paragraph after Eq.(**36**)) that the $l/X_m \ll 1$ condition is also fulfilled for smaller *rounded* grains (particles) with sizes at least up to $\sim 2\mu$m, i.e., up to the maximum clay particles. In other words, this assumption means that the single-valued dependence between $\Delta v_w$ and $X_m$ given by Eq.(**35**) also relates to rounded grains of size up to the maximum clay particles. We will show that the $l$ estimate flowing out of the above assumption is in the agreement with it and thereby justifies it.

Since, for the clay matrix (i.e., at $X_m <$ or $\sim 2\mu$m) $\Delta v_w >$ or $\sim 1$ [see the above point (ii)], the indicated assumption (the single-valued $\Delta v_w(X_m)$ dependence starting from silt-sand grains and up to the rounded grains with $X_m \sim 2\mu$m) means that for the rounded particles at $X_m \sim 2\mu$m one can write $\Delta v_w(X_m \sim 2\mu m) \sim 1$. This relation and Eq.(**35**) determine the $l$ value. However, before simple joint solving of these equations two specifications (or reservations) are necessary.

(a) The above assumption only relates to rounded clay particles. In connection with that one should note that clay particles differ from silt-sand grains not only in size, but in shape also (in statistical meaning). Sand and silt grains are characterized by some shape distribution (e.g., [18]), but as a whole they can be considered as rounded. Clay particles are, as a whole, plate-like. However, for our aims it is only important that many approximately rounded clay particles (including those of the maximum size) can be found among others. In estimating the $l$ value we use such clay particles of the maximum size.

(b) In connection with using Eq.(**35**) not only at silt-sand sizes, but also at the maximum clay particle size, it is important to remember that clay particles differ from the silt-sand grains not only in size and shape, but shrink-swell property as well. Equation (**35**) relates to the grain (particle) matrix in the water saturated state (see the derivation in Section **4.4** and Fig.(**4**)). The silt-sand grain size is retained, but the clay particle size increases with the increase in water content. In $\Delta v_w(X_m \sim 2\mu m) \sim 1$ for simplicity we used the maximum clay particle size in the oven-dried state ($\sim 2\mu$m). The corresponding maximum clay particle size in the water saturated state, $r_{mM}$ is usually in the range $\sim (3 \div 4)\mu$m depending on the $v_z$ value of the clay (see the estimate $r_{mM} \cong 2v_z^{-1/3}(\mu m)$ at the end of Section **2**; as a rule $v_z \sim (0.1 \div 0.4)$). In estimating the $l$ value below, we use the relation $\Delta v_w(X_m \sim 2\mu m) \sim 1$ with corrected $X_m \sim 3.5\mu$m as

$$\Delta v_w(X_m \sim 3.5\mu m) \sim 1 \ . \tag{37}$$

Equations (**35**) and (**37**) estimate the thickness of the additional water film, $l$ as

$$l \sim 0.3 \mu m \ . \tag{38}$$

One can be convinced that with the found *universal* (not depending on $X_m$) $l$ value the condition, $l/X_m \ll 1$ is fulfilled not only at $X_m > 20\mu$m, i.e., for silt and sand grains (as we assumed in Section **4.4**), but also at $X_m \sim 3.5\mu$m, i.e., for rounded clay particles of maximum size (as we assumed above in this Section). Thus, the above assumptions are justified and confirmed. Table **1** gives the estimates of the $\Delta v_w$ and $\Delta v_w/p$ values to illustrate the order of magnitudes at the typical $X_m$ and $p$ values of rigid grain matrices. These estimates seem to be quite reasonable since the physical conditions from Eq.(**22**) reflecting the specific features of rigid grain matrices are fulfilled. Below Eqs.(**35**) and (**38**) will be used in the analysis of available data at different $X_m$ values.

Thus, to calculate the $Q$ factor of the rigid grain matrix (Eq.(**27**)) one needs to know the usual sand parameters, the porosity, $p$ and maximum grain size, $X_m$ instead of the $v_s$ and $v_z$ parameters in the case of the $Q$ factor of the shrink-swell clay matrix. Indeed, $p$ and $X_m$ [plus the universal water film thickness, $l$ from Eq.(**38**)] permit the step-by-step estimation of $\Delta v_w$ [Eq.(**35**)], $v_s$ and $v_z$ [Eq.(**23**)], $v(\zeta)$ [Eq.(**26**)], and $Q$ [Eq.(**27**)] for "shrinking" sand formally using the approach that was earlier developed for clays. In addition, the solid density (of silt-sand grains or clay particles), $\rho_s$ is needed to transit to gravimetric water content in both cases of sand (Section **4.3**) or clay.

### 4.6. The *H* Factor of the Rigid-Grain Matrix

In Sections **4.1**-**4.5** we showed that the method that was proposed for finding the *Q* factor of clay (Section **2**) after some adaptation can be applied for finding the *Q* factor of sand. The key point of the adaptation is the consideration of the sand *Q* factor as originating from some specific "shrinkage" curve. The physical nature of the "shrinkage" (unlike the true clay matrix shrinkage) is connected with the additional water film of the characteristic minimum thickness between grains of preliminarily saturated matrix. After expressing the sand "shrinkage" curve parameters, $v_s$, $v_z$, $\zeta_z$ (Fig.(**3**)) through the usual parameters of the sand, the porosity, $p$ and maximum grain size $X_m$ as well as the universal characteristic thickness of the water film, $l$ (see Eq.(**23**), (**35**), and (**38**)) we can consider the sand to be a "clay" with some specific shrinkage features (Section **4.1**), and then, relying on the general formalism for clays (Section **2**) we can ascribe to the sand the "shrinkage" curve (Eq.(**26**)) and *Q* factor (Eq.(**27**)).

In light of that stated above, it is clear that after expressing the sand "shrinkage" parameters, $v_s$, $v_z$, $\zeta_z$, and the saturation degree, $F(\zeta)$ through $p$, $X_m$, and $l$, the general method for finding the *H* factor of clay (Section **2**), taking its modifications into account (see Sections **3.2**, **3.3**, and below), can also be used for finding the *H* factor of sand with some small adaptation.

Thus, we proceed from the sand parameters, $p$ and $X_m$ plus $l$ to find $v_s$, $v_z$, $\zeta_z$ (Eq.(**23**), (**35**), and (**38**)) and "shrinkage" curve (Eq.(**26**)) for the sand. The expression for the *H* factor of the "shrinking" sand and the range of its applicability are given by Eq.(**3**) because in this case the relative water content is in the $\zeta_* < \zeta \leq \zeta_o = 1$ range (for $\zeta_o$ see Fig.(**2**)), but not $\zeta_* < \zeta \leq \zeta_o = \zeta_h$ as for clay (after modification noted in Section **3.1**).

In the case of sand the characteristic size, $R(\zeta)$ (entering Eq.(**3**)) is presented as

$$R(\zeta) = \begin{cases} \rho'_m, & \zeta_z < \zeta \leq 1 \\ \rho'_c(\zeta), & \zeta_* \cong 0.1\zeta_z < \zeta \leq \zeta_z \end{cases}. \tag{39}$$

That is, in the case of sand the structure of the $R(\zeta)$ expression for clay (Eq.(**4**)) is kept, but with (i) replacement $\zeta_n \to \zeta_z$ (as noted after Eq.(**11**) $\zeta_z$ for sand plays part of both $\zeta_z$ and $\zeta_n$); (ii) $\rho'_m$=const; and (iii) quantitatively another $\rho'_c(\zeta)$ dependence. The similarity and difference between sand and clay is visually illustrated by comparison between Figs.(**5**) and (**1**). In particular, the range, $\zeta_n < \zeta \leq \zeta_h$ in Fig.(**1**) corresponds to $\zeta_z < \zeta \leq 1$ in Fig.(**5**), and curves 1, 2, 3, 4 correspond with each other. At $\zeta_z < \zeta \leq 1$ the maximum size of a pore-tube cross-section, $R(\zeta) = \rho'_m$=const (Eq.(**39**) and Fig.(**5**), curve 2) is connected with its 3D analogue (see below). To find the characteristic size of a pore-tube cross-section, $R(\zeta) = \rho'_c(\zeta)$ in the range $\zeta_* < \zeta \leq \zeta_z$ (Eq.(**39**); Fig.(**5**), curve 4), we should solve the same Eq.(**5**) that expresses the water balance in a sand cross-section. However, the $F(\zeta)$ and $\varphi(\rho')$ functions entering Eq.(**5**) in the case of sand are different than for clay.

The saturation degree of the sand, $F(\zeta)$ is given by Eq.(**6**) with the "shrinkage" curve, $v(\zeta)$ from Eq.(**26**). One can check that for sand (unlike clays, see Eq.(**6**))

$$F(\zeta) = \zeta/\zeta_z, \qquad 0 < \zeta \leq \zeta_z \tag{40}$$

[use Eq.(**11**)] and $F(\zeta)$=1 at $\zeta_z < \zeta \leq 1$. This expression for $F(\zeta)$ of sand can be written directly from definitions of $\zeta$ and $\zeta_z$. Obtaining it from Eq.(**6**) and (**26**) gives additional evidence that the sand "shrinkage" curve (Eq.(**26**)) is reasonable.

For sand we use the simplest distribution, $\varphi(x(\rho'))$ for the two-dimensional situation from the intersecting-surfaces approach [18] as

$$\varphi(x(\rho')) = [1 - (1-p)^{I(x(\rho'))/I(1)}]/p, \tag{41}$$

$$x(\rho') = (\rho' - \rho'_{min})/(\rho'_m - \rho'_{min}), \qquad 0 < x \leq 1; \tag{42}$$

$$I(x)=\ln(6)(3x)^3\exp(-3x), \qquad (I(1)=2.4086) \qquad 0<x\leq 1; \tag{43}$$

with *constant* values of the minimum ($\rho'_{min}$) and maximum ($\rho'_m$) pore-tube cross-section sizes (unlike clay case [11]) and the *constant* sand porosity, $p$ at $0\leq\zeta\leq\zeta_z$ (unlike the varying clay porosity [14]).

When solving Eq.(**5**) with $F(\zeta)$ from Eq.(**40**) and $\varphi(\rho')$ from Eqs.(**41**)-(**43**), we use the following boundary conditions.

(i) At the boundary water content $\zeta=\zeta_*$ the maximum internal size of the water filled ($\rho'_f(\zeta)$; Fig.(**5**), curve 3) and water containing ($\rho'_c(\zeta)$; Fig.(**5**), curve 4) pore-tube cross-sections should coincide (this condition was not used in [11])

$$\rho'_f(\zeta_*)=\rho'_c(\zeta_*) . \tag{44}$$

(ii) At the boundary water content $\zeta=\zeta_z$ $\rho'_f(\zeta)$ (Fig.(**5**), curve 3) and $\rho'_c(\zeta)$ (Fig.(**5**), curve 4) also coincide (Fig.(**5**))

$$\rho'_f(\zeta_z)=\rho'_c(\zeta_z) . \tag{45}$$

(iii) $R(\zeta)$ (Eq.(**39**)) should be smooth at $\zeta=\zeta_z$ (Fig.(**5**)). That is,

$$\rho'_c(\zeta_z)=\rho'_m \qquad \text{and} \qquad d\rho'_c(\zeta)/d\zeta\big|_{\zeta=\zeta_z}=0 . \tag{46}$$

In addition, the $\rho'_f(\zeta)$ and $\rho'_c(\zeta)$ functions (Fig.(**5**)) should meet the obvious physical condition (which was not used in [11]) that the water-containing (i.e., non-totally filled) pore tubes give a small contribution to the water balance equation (Eq.(**5**)). That is, in Eq.(**5**)

$$\varphi(\rho'_f) \gg \int_{\rho'_f}^{\rho'_c} g(\rho')\frac{d\varphi}{d\rho'}d\rho' . \tag{47}$$

It follows that independently of an exact form of $g(\rho')$ dependence, $\rho'_c(\zeta)$ differs from $\rho'_f(\zeta)$ by the small addition of $\delta\rho'_f$ as (Fig.(**5**))

$$\rho'_c=\rho'_f+\delta\rho'_f , \qquad \delta\rho'_f(\zeta)\ll\rho'_f(\zeta), \qquad \zeta_*<\zeta<\zeta_z \tag{48}$$

and according Eq.(**44**) and (**45**)

$$\delta\rho'_f(\zeta_*)=\delta\rho'_f(\zeta_z)=0 . \tag{49}$$

Then, without additional assumptions with respect to the $g(\rho')$ function and pore shape, we can find $\rho'_c(\zeta)$ in the range $\zeta_*<\zeta<\zeta_z$ in the first approximation as $\rho'_c(\zeta)=\rho'_f(\zeta)$ where $\rho'_f(\zeta)$ (Fig.(**5**), curve 3) is the solution of the equation

$$F(\zeta)=\varphi(\rho'_f) . \tag{50}$$

Solving Eq.(**50**) with $F(\zeta)$ from Eq.(**40**) and $\varphi(x)$ from Eqs.(**41**) and (**43**), one finds $x(\zeta)$ at $\zeta_*<\zeta\leq\zeta_z$. Then, equalizing the $x(\zeta)$ found to $x(\rho')$ from Eq.(**42**) at given $\rho'_m$ and $\rho'_{min}$, one finds from $x(\zeta)=x(\rho')$ the $\rho'_f(\zeta)$ dependence at $\zeta_*<\zeta\leq\zeta_z$.

In the second approximation one can write $\rho'_c(\zeta)$ at $\zeta_*<\zeta\leq\zeta_z$ (Fig.(**5**), curve 4) as

$$\rho'_c(\zeta) = \begin{cases} \rho'_f(\zeta), & \zeta_*<\zeta\leq\zeta' \\ \rho'_f(\zeta)+\delta\rho'_f(\zeta), & \zeta'<\zeta\leq\zeta_z \end{cases} . \tag{51}$$

It follows from the general qualitative picture (Fig.(**5**)) that max($\delta\rho'_f$)=max($\rho'_c-\rho'_f$) is reached close to $\zeta=\zeta_z$. That is, the point of "sewing" $\zeta=\zeta'$ (Fig.(**5**)) is also close to $\zeta=\zeta_z$ (i.e., $\zeta_z-\zeta'<<\zeta'-\zeta_*$; that is confirmed by direct calculations in Section **5**). For this reason we approximate $\rho'_c=\rho'_f+\delta\rho'_f$ at $\zeta'<\zeta\leq\zeta_z$ to be

$$\rho'_c(\zeta)/\rho'_m = 1 + D(\zeta-\zeta_z)^2 , \qquad \zeta'<\zeta\leq\zeta_z . \tag{52}$$

Then conditions at $\zeta=\zeta_z$ (Fig.(**5**)) given by Eq.(**46**) are obviously fulfilled. The $D$ coefficient and "sewing" point $\zeta=\zeta'$ (Fig.(**5**)) are found from conditions of the smooth connection between $\rho'_c(\zeta)=\rho'_f(\zeta)$ from Eq.(**51**) and $\rho'_c(\zeta)$ from Eq.(**52**) at $\zeta=\zeta'$. The $\rho'_c(\zeta)$ found in good approximation meets Eq.(**5**) and conditions from Eq.(**44**)-(**47**) at $\zeta_*<\zeta\leq\zeta_z$ (Fig.(**5**)).

Finally, calculation of $R(\zeta)$ (Eq.(**39**); Fig.(**5**), curves 2 and 4) together with Eq.(**3**) gives $H(\zeta)$. The $Q(\zeta)$ factor from Eq.(**27**) and this $H(\zeta)$ give the sand-water retention curve, $h(\zeta)=H(\zeta)Q(\zeta)$. Replacement of $\zeta$ with $W/W_M$ (Eqs.(**32**) and (**28**)) or recalculation of $\zeta$ to $\theta$ (Eq.(**34**) with $v(\zeta)$ from Eq.(**26**)) enables transition to the customary gravimetric water content, $W$ or volumetric water content, $\theta$ (in particular, $\theta'$ corresponds to $\zeta=\zeta'$ and $v=v(\zeta')$ in Eq.(**34**)).

The minimum ($\rho'_{min}$) and maximum ($\rho'_m$) internal sizes of pore-tube cross-sections in sand, that were used above (see Eqs.(**39**) and (**42**)), can be expressed through the minimum ($r'_{min}$) and maximum ($r'_m$) (three-dimensional) internal pore sizes as $\rho'_{min}=(3/4)r'_{min}$ and $\rho'_m=(3/4)r'_m$ [14, 11]. Thus, the sand-water retention, in general, is determined by the following parameters, $\rho_s$, $p$, $X_m$, $r'_m$ and $r'_{min}$. When using the volumetric water content one only needs input data on $p$, $X_m$, $r'_m$ and $r'_{min}$.

## 5. DATA AND THEIR ANALYSIS

The important requirement to possible data is the presence of all the natural sand fractions from $X_m$ to $X_{min}\sim 20$-$50\mu m$ ($X_{min}$ is the minimum sand grain size). Otherwise, the estimate $\Delta v_w=12l/X_m$ (Eq.(**35**)) is non-applicable (because $d\neq X_m/4$).

As noted, to obtain the model prediction one needs input data on $p$, $X_m$, $r'_m$ and $r'_{min}$. To estimate the $r'_m$ and $r'_{min}$ values we used relations $r'_m\cong 0.15X_m$ and $r'_{min}\cong 0.15X_{min}$ that are explained in Fig.(**6**). Thus, input data are only reduced to the $p$, $X_m$, and $X_{min}$ values (usually $X_{min}\sim 20$-$50\mu m$). We used the available suitable drainage data of four sandy soils: one sand from Lamara and Derriche [21] and three sands from Elrick et al. [22] (quoted by [2]). Data on the sand porosity ($p$) and the minimum grain size ($X_{min}$) for the four sands are indicated in Figs.(**7**)-(**10**) and Table **2**. The $X_m$ value that was extracted from [21] is indicated in Fig.(**7**) and Table **2**. Haverkamp and Parlange [2] do not give a clear indication on $X_m$ for soils presented in Figs.(**8**)-(**10**). However, it is obvious that in any case $X_m\sim 1000$-$2000\mu m$. The $X_m$ values for the corresponding three sands were estimated in the data analysis and also indicated in Figs.(**8**)-(**10**) and Table **2**. They are in the above range (for these $X_m$ values see Section **6**).

The $p$, $X_m$, and $X_{min}$ data (Table **2**) were analyzed according to the two-factor model of sand water retention (Section **4**). It was natural to present the model predicted water retention curves in Figs.(**7**)-(**10**) (solid lines) in the same units as the corresponding experimental data and fitted curves from [21] and [22] (quoted by [2]) (Figs.(**7**)-(**10**); white circles). Then, we compared the model predicted water retention curves with the experimental sand-water retention curves as well as with the fitted water retention curves from [21] and [2] (Figs.(**7**)-(**10**); dashed lines) and with the bands of the maximum probable error of the fitted curves from [2] (Figs.(**8**)-(**10**); dotted lines).

## 6. RESULTS AND DISCUSSION

The results presented in Figs.(**7**)-(**10**) (solid lines) and Table **2** clearly show that the sand water retention curves predicted by the two-factor model developed in Section **4** are in the good agreement with the experimental data and the fitted curves found in [21] and [2] as well as with the bands of the maximum probable error of the fitted curves [2] in Figs.(**8**)-(**10**). In Figs.(**8**)-(**10**), in fact, we used $X_m$ as a preliminarily estimated parameter with accuracy $\sim\pm 50\mu m$ for lack of reliable data. However, evaluating the good agreement found above one should take into account the following points: (i) the agreement between prediction and data in Fig.(**7**) where $X_m$ is a given experimental value (but not a preliminarily estimated one) emphasizes the feasibility of the model as applied to the prediction of the sand (or rigid matrix) water retention; (ii) the agreement between the non-fitted prediction (solid line) and curve fitted with Fredlund and Xing's [23] fitting model in Fig.(**7**) (dashed line) as well as between the prediction and curve fitted with van Genuchten's [1] fitting model (not shown in Fig.(**7**); cf. Fig.(**7**) with Fig.(**3**) in Lamara and Derriche [21]) additionally underlines the feasibility of the model; (iii) $X_m$ has the clear and simple physical meaning and is an obviously measurable parameter; (iv) the preliminarily estimated $X_m$ values (Figs.(**8**)-(**10**)) are in the reasonable range 1000µm<$X_m$<2000µm; and, finally, (v) we were limited in Figs.(**8**)-(**10**) to only one preliminarily estimated parameter ($X_m$) (unlike several such parameters in [21] and [2] as well as in other models indicated in Section 1). In any case the predicted curves are within the limits of experimental errors.

The model estimated parameters in Table **2**, in addition, illustrate that for the real sandy soils as noted above, $\theta_z$ and $\theta_M$ are very close each other and $\Delta v_w \ll 1$ (cf. Table **1**). In addition, the "sewing" point $\theta'$ is less than $\theta_z$, as it should be, but very close to $\theta_z$ for all the four sands and for this reason is not given in Table **2**.

In this work we considered the simplest version of the two-factor model as applied to sand water retention (Section **4**). The model can be specified and improved (keeping its physical, but not fitting character) as follows.

(i) One can consider the two-factor model of sand water retention in the following, second (square) approximation with respect to the small $\Delta v_w$ value. This approximation can additionally soften the sharp bend of the sand water retention curve near saturation (see Figs.(**2**) and (**7**)-(**10**)), and be relevant for finer sands (i.e., sufficiently small $X_m$).

(ii) For the more accurate water retention description of many real sands one can use the two-mode $\varphi(\rho')$ distribution from [18] instead of Eqs.(**41**)-(**43**).

(iii) The prediction accuracy of the sand water retention at small water contents, near $\theta = \theta_*$ (see Figs.(**7**)-(**10**)) (in the vicinity of the model applicability boundary, $\theta_*$) can be improved accounting for some possible internal transformation of the initial $\varphi$ grain size distribution in the range of the smallest grains at the expense of the self-comminution of the larger contacting grains.

Finally, it is worth noting the possibility of the simple adaptation of the above model and results to the more complex, but related case of the matrix consisting of saturated aggregates. When dewatering the inter-aggregate pores of an aggregated soil (if they are water filled), aggregates remain saturated and should behave similar to the rigid grain system. If the inter-aggregate pores are capillary ones the corresponding possible small "tail" of the main soil water retention curve (this tail is connected with the loss of inter-aggregate water), can be considered using the above model after replacing the inter-grain pores of a sand with inter-aggregate ones of the soil. Unfortunately, reliable data for such consideration of the "tail" are lacking at present.

## 7. CONCLUSION

The aim of this work is to propose an approach to the water retention of rigid soils that, unlike the models of the curve-fitting type, could be useful not only for engineering applications, but also for improving our knowledge and understanding. The novelty points are as follows. (1) The general analysis of the observed soil water retention curves in the frame of the two-factor model of clay water retention, enabled us to formally ascribe some "shrinkage" curve to any soil, including a soil with a rigid grain matrix. (2) As applied to the rigid soil we use the term *pseudo shrinkage* because its nature has no relation to true shrinkage of clay or clay soil. The analysis of the number of *specific features* of a rigid matrix that flow out of the *simple generally known facts*, permitted us to connect this very weak "shrinkage" of rigid soil with an additional water volume that is distributed in the already saturated rigid matrix as a *supplementary water film along the grain surfaces*. (3) This additional water film (with the thickness ~0.3µm) is associated with the closest vicinity of the zero suction point and is quickly lost with rigid soil dewatering and the rapid (nearly vertical) ascent of soil suction (in shrink-swell soils the similar ascent is rather smoother owing to the "work" of real shrinkage). (4) The pseudo shrinkage of rigid soils allows one to consider the soils as "clay" and use the available method of the two-factor model of clay water retention, after some preliminary expression of the corresponding parameters of such (pseudo) clay through the usual parameters of the rigid grain matrix. (5) Eventual input parameters of the developed model of rigid-soil water retention have a clear physical meaning, are simply measured, and include the porosity ($p$) of a rigid grain system as well as the maximum ($X_m$) and minimum ($X_{min}$) sizes of grains.

The *major result* is the promising agreement between predicted and experimental water retention curves of the four sandy soils. The *principle* result is the confirmed fruitfulness of the two-factor model not only for a shrinking clay, but also for a rigid soil. This suggests the possibility of developing the similar model in the general case of an aggregated soil.

## NOTATION

*A*    characteristic constant of a clay microstructure [14], dimensionless
*D*    coefficient in Eq.(**52**), dimensionless
*d*    mean grain size, µm

| | |
|---|---|
| $F(\zeta)$ | saturation degree, dimensionless |
| $g$ | specific gravity, m s$^{-2}$ |
| $g(\rho')$ | filling degree of the pore tubes of internal $\rho'$ size with water, dimensionless |
| $H$ | factor originating from the adsorption-capillary phenomena, kPa or cm of H$_2$O |
| $h$ | soil suction, kPa or cm of H$_2$O |
| $I(x)$ | function from Eq.(**43**), dimensionless |
| $k$ | coefficient in the "shrinkage" curve of the sand, dimensionless |
| $L$ | side size of sand volume of cubic shape, m |
| $l$ | characteristic minimum thickness of the additional water film between the grains, μm |
| $l_o$ | thickness of a monomolecular water layer, ~3Å |
| $P(W)$ | soil porosity, dimensionless |
| $p$ | constant sand porosity in the area $0<\zeta\leq\zeta_z$, dimensionless |
| $Q$ | factor originating from the shrinkage-swelling of the soil matrix, dimensionless |
| $R$ | characteristic size of pore-tube cross-section of clay matrix or sand, μm |
| $r$ | external pore size that includes a half-thickness of clay particles limiting the pores, μm |
| $r_{mM}$ | maximum external size of clay pores at $\zeta=1$, μm |
| $r_{mz}$ | maximum size of clay particles in oven-dried state, μm |
| $r_o$ | minimum pore size of the solid particle system, μm |
| $r_{oM}$ | minimum size of the water saturated pores in clay, μm |
| $r'$ | internal pore size, μm |
| $r'_m$ | maximum (three-dimensional) internal pore size, μm |
| $r'_{min}$ | minimum (three-dimensional) internal pore size, μm |
| $V_M$ | maximum volume of "shrinking" sand with the minimum water film at $\zeta=1$, m$^3$ |
| $V_p$ | pore volume of sand in the area of the rigid matrix, $0<\zeta\leq\zeta_z$, m$^3$ |
| $V_s$ | volume of sand solid phase, m$^3$ |
| $V_z$ | minimum volume of "shrinking" sand when its matrix is rigid, $0<\zeta\leq\zeta_z$, m$^3$ |
| $v(\zeta)$ | shrinkage curve of clay in terms of relative volume (the ratio of the specific volume to that at the liquid limit of the clay); "shrinkage" curve of sand, dimensionless |
| $v_s$ | relative volume of clay solids (the ratio of the solid volume to clay volume at the liquid limit); corresponding parameter of "shrinking" sand, dimensionless |
| $v_z$ | $v$ value at the shrinkage limit of clay; corresponding parameter of "shrinking" sand, dimensionless |
| $W$ | gravimetric water content of sand, g g$^{-1}$ |
| $W_M$ | maximum value of $W$ at $\zeta=1$, g g$^{-1}$ |
| $w$ | gravimetric water content of clay, g g$^{-1}$ |
| $w_M$ | liquid limit of clay, g g$^{-1}$ |
| $w_h$ | maximum swelling point of clay, g g$^{-1}$ |
| $w_*$ | gravimetric water content of clay at $\zeta=\zeta_*$, gg$^{-1}$ |
| $X_m$ | characteristic (maximum) sand grain size, μm |
| $X_{min}$ | minimum sand grain size, μm |
| $x, y, z$ | orthogonal axes in the sand cube, m |
| $x(\rho')$ | parameter from Eq.(**42**), dimensionless |
| $\alpha$ | characteristic constant of a clay microstructure [14], dimensionless |
| $\alpha_c$ | contact angle, degrees |
| $\Gamma$ | surface tension of water, N m$^{-1}$ |
| $\Delta V$ | increment of "shrinking" sand volume between $V_M$ and $V_z$, m$^3$ |
| $\Delta V_w$ | increment of water volume associated with sand volume increment $\Delta V$, m$^3$ |
| $\Delta v$ | relative additional sand volume (per unit volume of dry sand), dimensionless |
| $\Delta v_w$ | relative additional water volume (per unit volume of dry sand), dimensionless |
| $\delta\rho'_f$ | small difference between $\rho'_c(\zeta)$ and $\rho'_f(\zeta)$, dimensionless |
| $\zeta$ | relative water content of clay (the ratio of the gravimetric water content, $w$ to the liquid limit of the clay, $w_M$) or sand, dimensionless |
| $\zeta_M$ | $\zeta=1$ value at clay liquid limit or sand zero suction, dimensionless |
| $\zeta_h$ | $\zeta$ value at maximum swelling point of clay, dimensionless |
| $\zeta_n$ | clay air-entry point, dimensionless |
| $\zeta_z$ | shrinkage limit of clay or "shrinking" sand, dimensionless |
| $\zeta_o$ | zero suction point; for clay $\zeta_o\equiv\zeta_h$; for sand $\zeta_o=1$, dimensionless |
| $\zeta_*$ | estimate of the range $\zeta_*<\zeta<1$ boundary where $H$ is only connected with capillarity, dimensionless |
| $\zeta'$ | $\zeta$ value where $\delta\rho'_f(\zeta')=0$, dimensionless |
| $\theta$ | volumetric water content of sand, cm$^3$ cm$^{-3}$ |
| $\theta_M$ | value of $\theta$ at $\zeta=1$, cm$^3$ cm$^{-3}$ |
| $\theta_z$ | value of $\theta$ at $\zeta=\zeta_z$, cm$^3$ cm$^{-3}$ |
| $\theta_*$ | value of $\theta$ at $\zeta=\zeta_*$, cm$^3$ cm$^{-3}$ |
| $\lambda$ | function of the ratios of $l_o$ to $X_m$ and $[2\Gamma/(\rho_w g)]^{1/2}$, dimensionless |
| $\lambda_o$ | value of $\lambda$ at $l_o/X_m=0$ and $l_o/[2\Gamma/(\rho_w g)]^{1/2}=0$, dimensionless |
| $\rho$ | external pore-tube size that includes a half-thickness of clay particles limiting the pores, µm |
| $\rho_s$ | sand solid density, g cm$^{-3}$ |
| $\rho_w$ | water density, g cm$^{-3}$ |
| $\rho'$ | internal pore-tube size, µm |
| $\rho'_c(\zeta)$ | maximum internal size of the water-containing pore tubes, µm |
| $\rho'_f(\zeta)$ | maximum internal size of the water-filled pore tubes, µm |
| $\rho'_m(\zeta)$ | maximum internal size of pore tube cross-sections (constant for sand), µm |
| $\rho'_{min}$ | minimum internal size of pore tube cross-sections (constant for sand), µm |
| $\varphi(\rho')$ | pore tube-size distribution, dimensionless |

**Figure Captions**

**Fig.(1)**. Qualitative view of relative characteristic internal pore-tube cross-section sizes of a clay matrix against the relative water content (the modified Fig.(**4**) from [11]). "Relative" size means the ratio of a size to $r_{mM}$ (the maximum pore size at the liquid limit); subscript $i$ of $\rho'_i$ corresponds to the index of the shown curves, $i=1,\ldots,4$. 1-the maximum internal size of pore-tube cross-sections, $\rho'_m(\zeta)/r_{mM}$ at $0<\zeta<\zeta_n$: 2-the same size as on curve 1, but at $\zeta_n<\zeta<1$; 3-the maximum internal size of water-filled pore-tube cross-sections, $\rho'_f(\zeta)/r_{mM}$ at $\zeta_*<\zeta<\zeta_n$; 4-the maximum internal size of water-containing pore-tube cross-sections, $\rho'_c(\zeta)/r_{mM}$ at $\zeta_*<\zeta<\zeta_n$. The smooth curve composed of curve 2 at $\zeta_n<\zeta<1$ and curve 4 at $\zeta_*<\zeta<\zeta_n$ gives the relative characteristic size, $R(\zeta)/r_{mM}$ that determines the capillary factor $H$ as a function of the relative water content. $\zeta_*$, $\zeta_z$, $\zeta'$, $\zeta_n$, and $\zeta_h$ are relative water contents corresponding to the lower boundary of the model applicability, shrinkage limit, point where $\rho'_f(\zeta')=\rho'_c(\zeta')$, air-entry point, and maximum swelling point, respectively.

**Fig.(2)**. The general view of the $Q$ factor and relative $H$ factor of a swell-shrink clay and rigid sand matrix. $\zeta_z$ and $\zeta_o$ are characteristic points.

**Fig.(3)**. The illustrative "shrinkage" curve of a sand matrix in the relative coordinates at $\zeta_z$ and $v_z$ close to unity.

**Fig.(4)**. Schematic illustration for estimating the relative additional water volume, $\Delta v_w$ of a rigid grain matrix. (a) Three different layer systems, normal to axes $x$, $y$, $z$; $L$ is the size of the sand cube; $d=X_m/4$ is the mean grain size. (b) Part of the cross-section, normal to a layer system in the water saturated state. (c) The same situation, but in the state when the water films of the minimum characteristic thickness, $l(<<X_m)$ are added along the boundaries of the grain layers, moving them apart.

**Fig.(5)**. Qualitative view of relative characteristic pore-tube cross-section sizes of a sand matrix against the relative water content. "Relative" size means the ratio of a size to $r'_m$; subscript $i$ of $\rho'_i$ corresponds to the index of the shown curves, $i=1,\ldots,4$. 1-the maximum size of pore-tube cross-sections, $\rho'_m/r'_m$ at $0<\zeta<\zeta_z$: 2-the same size as on curve 1, but at $\zeta_z<\zeta<1$; 3-the maximum size of water-filled pore-tube cross-sections, $\rho'_f(\zeta)/r'_m$ at $\zeta_*<\zeta<\zeta_z$; 4-the maximum size of water-containing pore-tube cross-sections, $\rho'_c(\zeta)/r'_m$ at $\zeta_*<\zeta<\zeta_z$. The smooth curve composed of curve 2 at $\zeta_z<\zeta<1$ and curve 4 at $\zeta_*<\zeta<\zeta_z$ gives the relative characteristic size, $R(\zeta)/r'_m$ that determines the capillary factor $H$ as a function of the relative water content.

**Fig.(6)**. The approximate estimate of the maximum ($r'_m$) or minimum ($r'_{min}$) internal size, of the inter-grain pores (white circle of BC radius) at the maximum ($X_m$) or minimum ($X_{min}$) size of grains (grey circles of AB radius). One successively has: (i) AE=$X_m$; (ii) AD=($\sqrt{3}/2$)AE=($\sqrt{3}/2$)$X_m$; (iii) AC=(2/3)AD=($\sqrt{3}/3$)$X_m$; (iv) AB=AE/2=$X_m/2$; (v) $r'_m$=2BC=2(AC-AB)=2($\sqrt{3}/3-1/2$)$X_m \cong 0.15X_m$. After replacing the above $X_m$ with $X_{min}$ and $r'_m$ with $r'_{min}$ one obtains $r'_{min} \cong 0.15X_{min}$.

**Fig.(7)**. The water retention (drying branch) curve data (white circles) of a dune sand [21], and the curve (solid line) that was predicted (without fitting) by the two-factor model from the data on the sand porosity ($p$=0.321) as well as maximum and minimum grain sizes ($X_m$=900μm and $X_{min}$=45μm) (the data from [21]). The black circles indicate the lower boundary of model applicability ($\theta_*$) and water content of the saturated sand ($\theta_z$). $\theta_M$ is the maximum water content (close to $\theta_z$) that corresponds to the zero suction. The dashed line gives the fitted curve [21] of the data (white circles) that was found from Fredlund and Xing's [23] fitting model.

**Fig.(8)**. The water retention (drying branch) curve data (white circles) of Preston sand from Elrick et al. [22] (quoted by Haverkamp and Parlange [2]), and the curve (solid line) that was predicted (without fitting) by the two-factor model using the data on the sand porosity ($p$=0.393) from [22]) (quoted by [2]) as well as maximum and minimum grain size, $X_m$=1200μm and $X_{min}$=50μm. The meaning of black circles is as in Fig.(**7**). The dashed line gives the fitted curve [2] to the data (white circles). Two dotted lines determine the band [2] that corresponds to the maximum probable error of the fitted curve (dashed line).

**Fig.(9)**. As in Fig.(**8**), but for Gormley sand ($p$=0.357, $X_m$=1700μm, $X_{min}$=50μm).

**Fig.(10)**. As in Fig.(**8**), but for Bolton sand ($p$=0.373, $X_m$=1200μm, $X_{min}$=50μm).

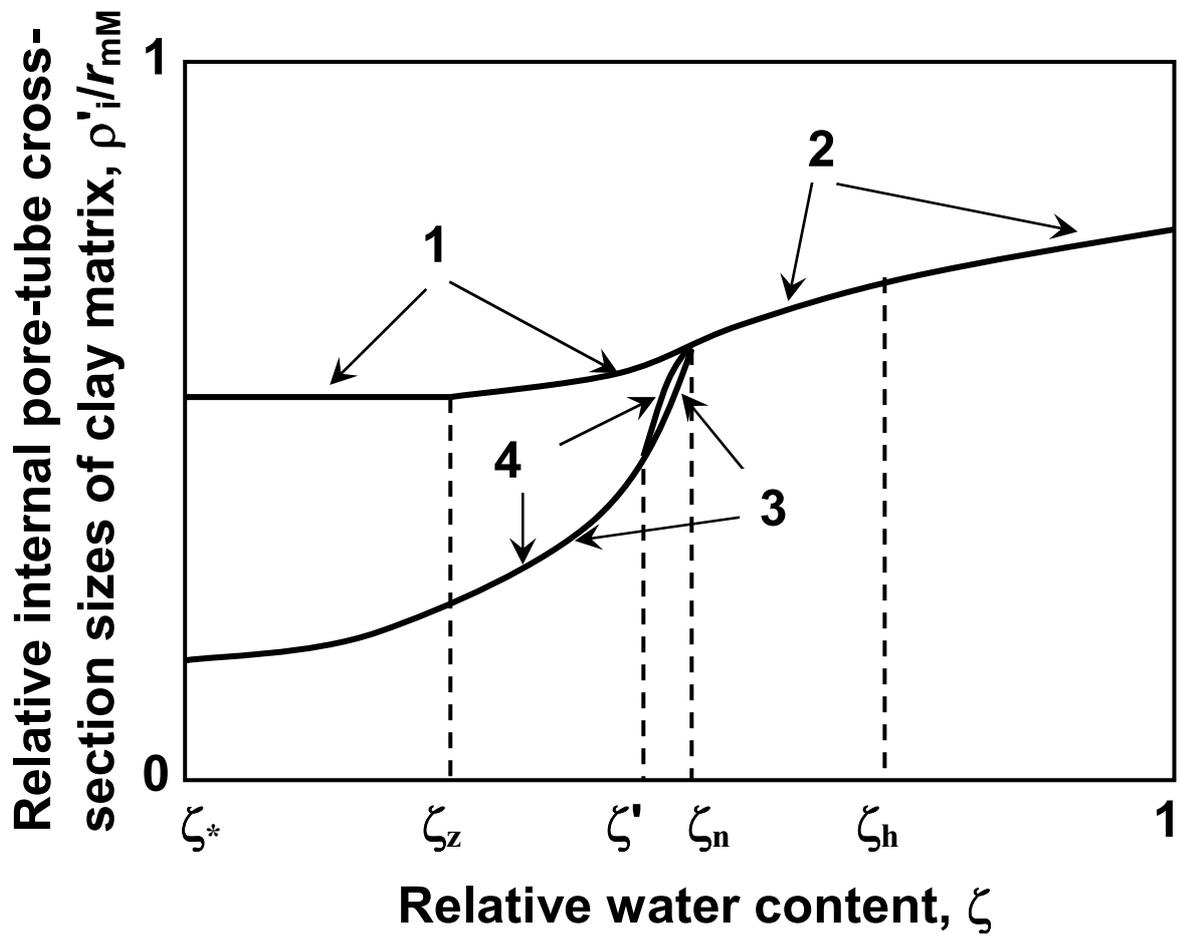

Fig.1

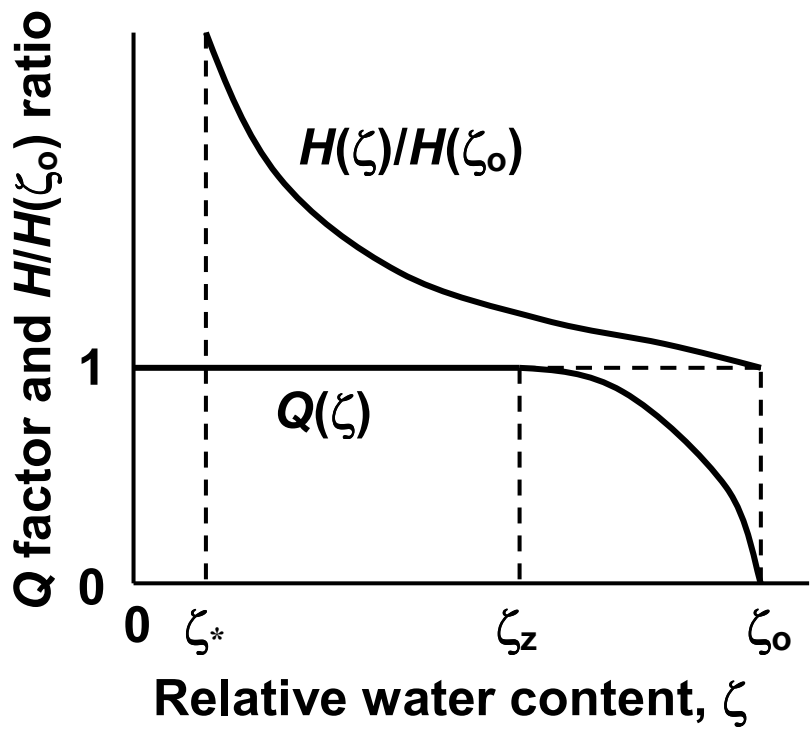

Fig.2

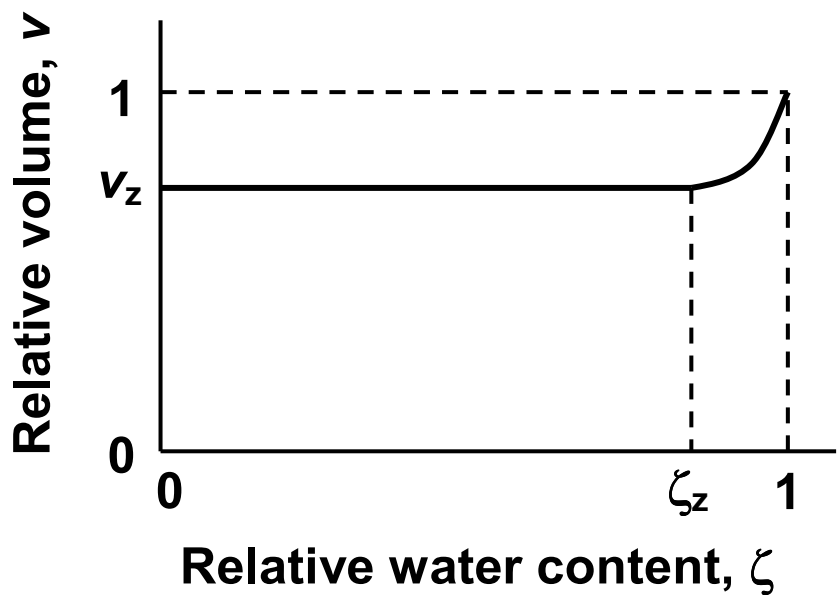

Fig.3

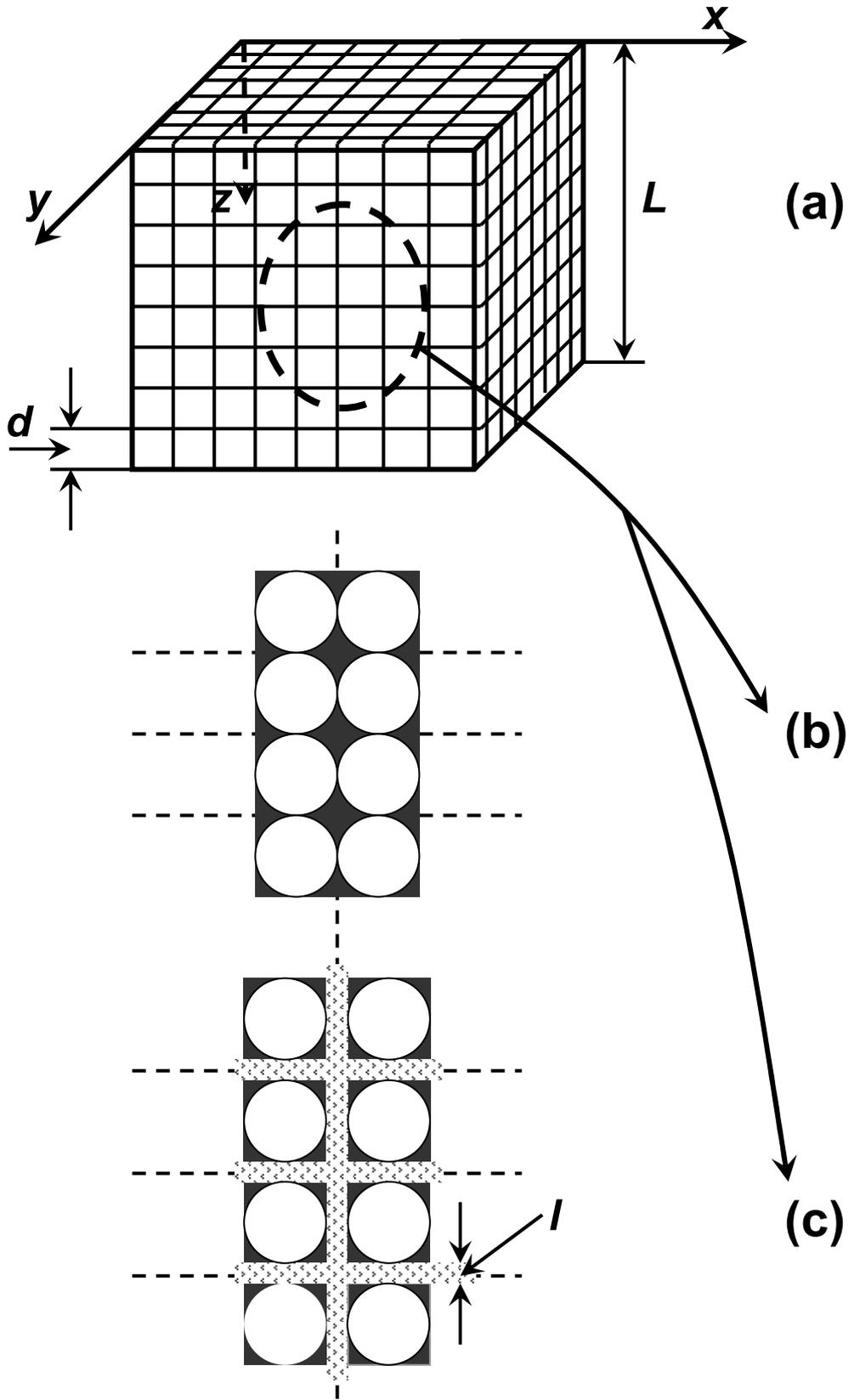

Fig.4

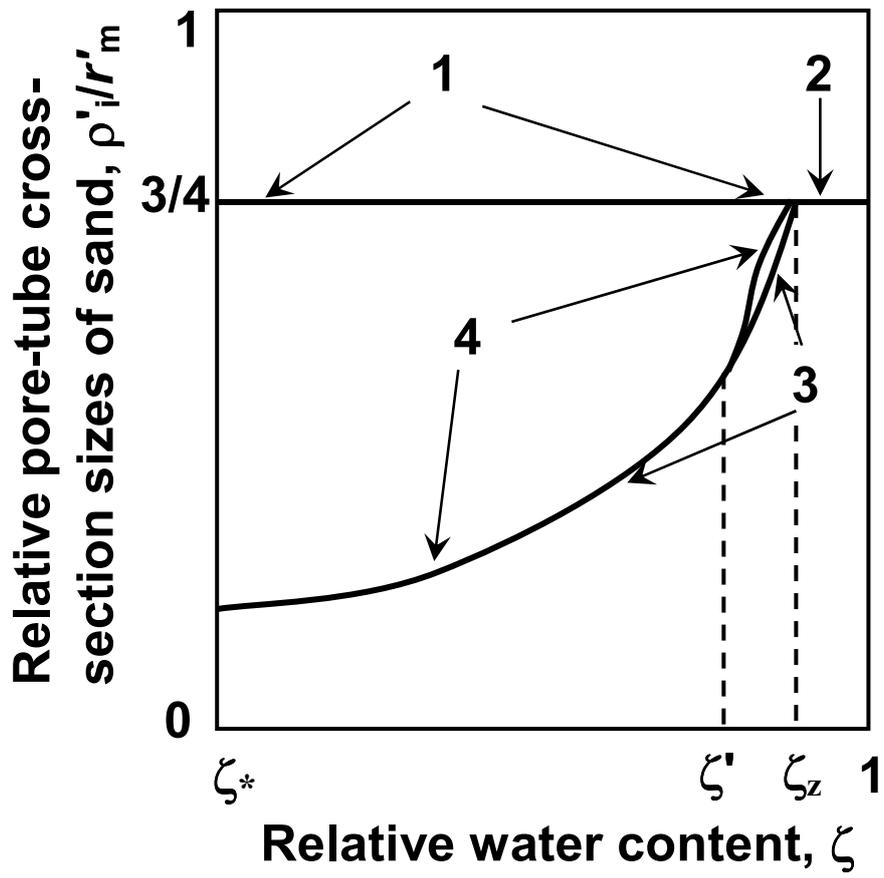

Fig.5

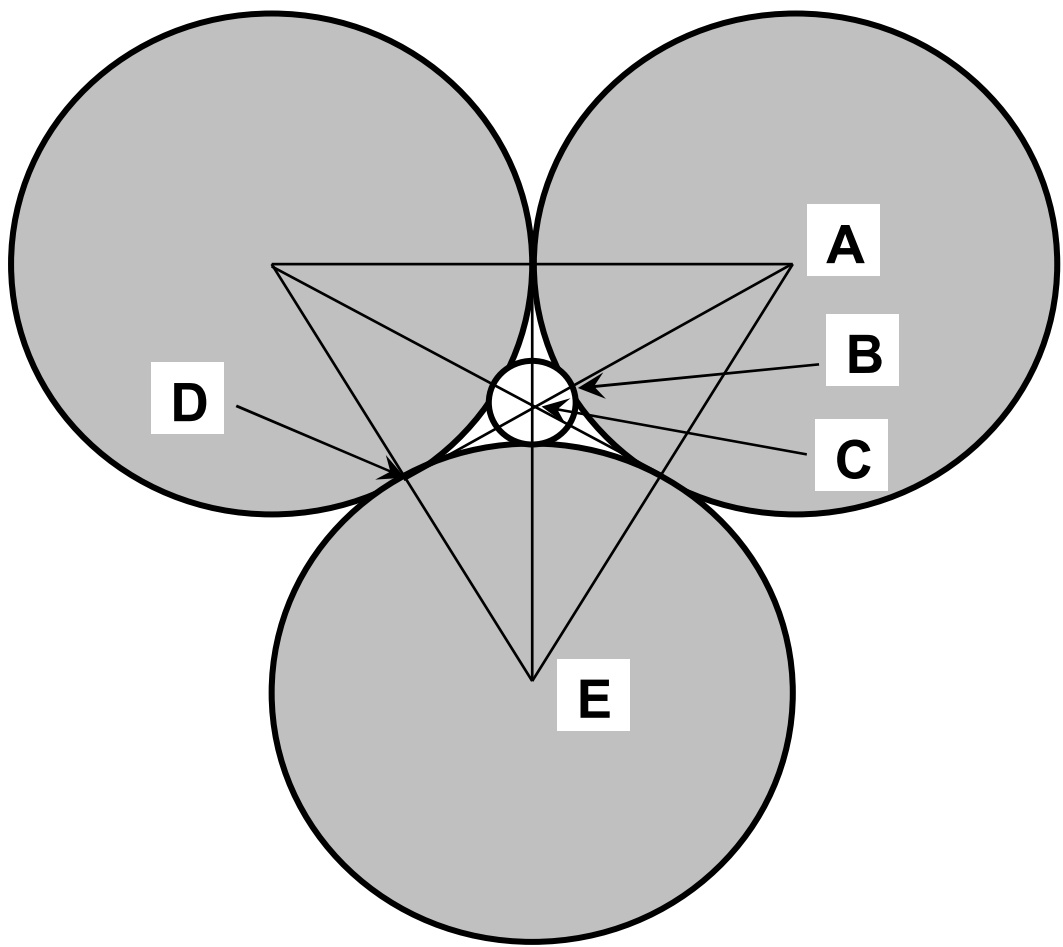

Fig.6

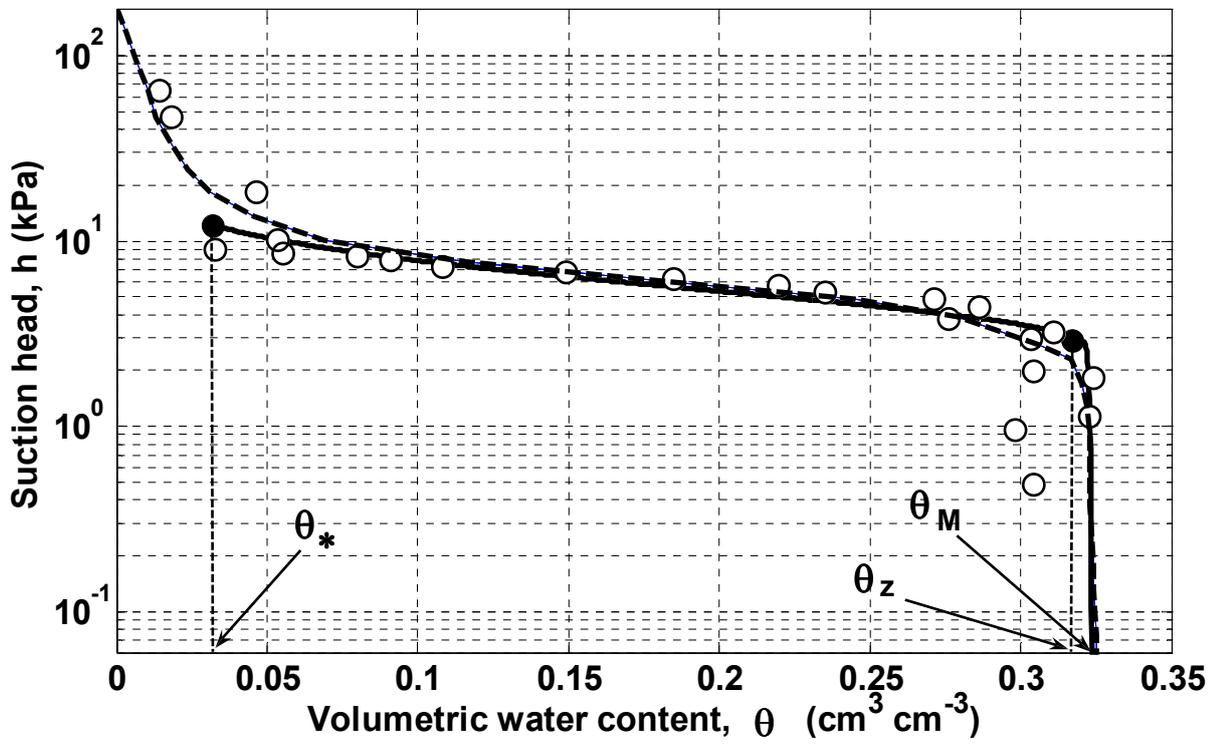

Fig.7

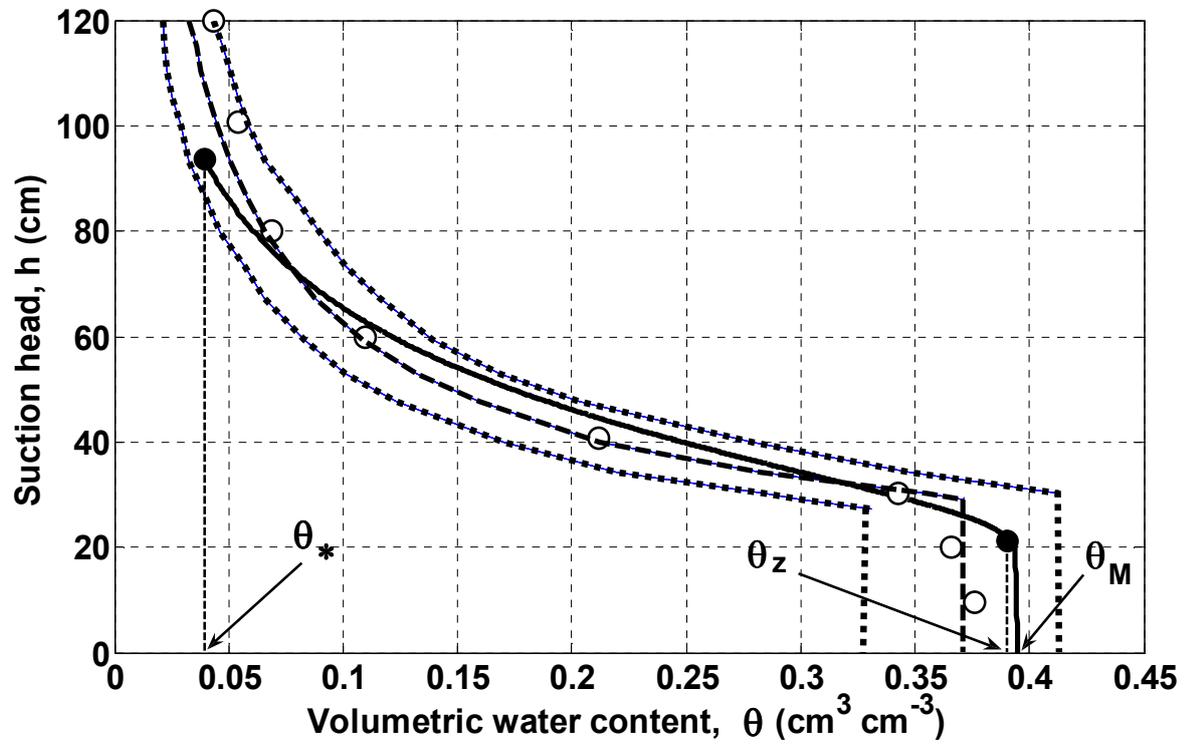

Fig.8

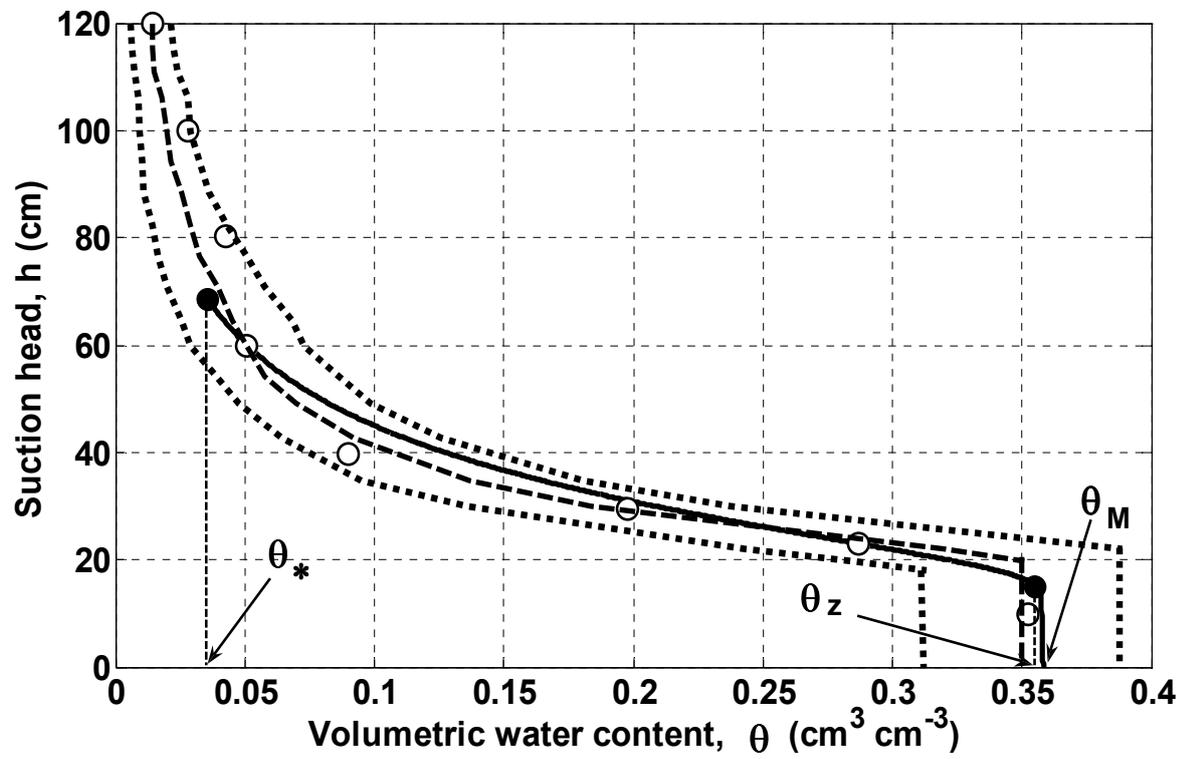

Fig.9

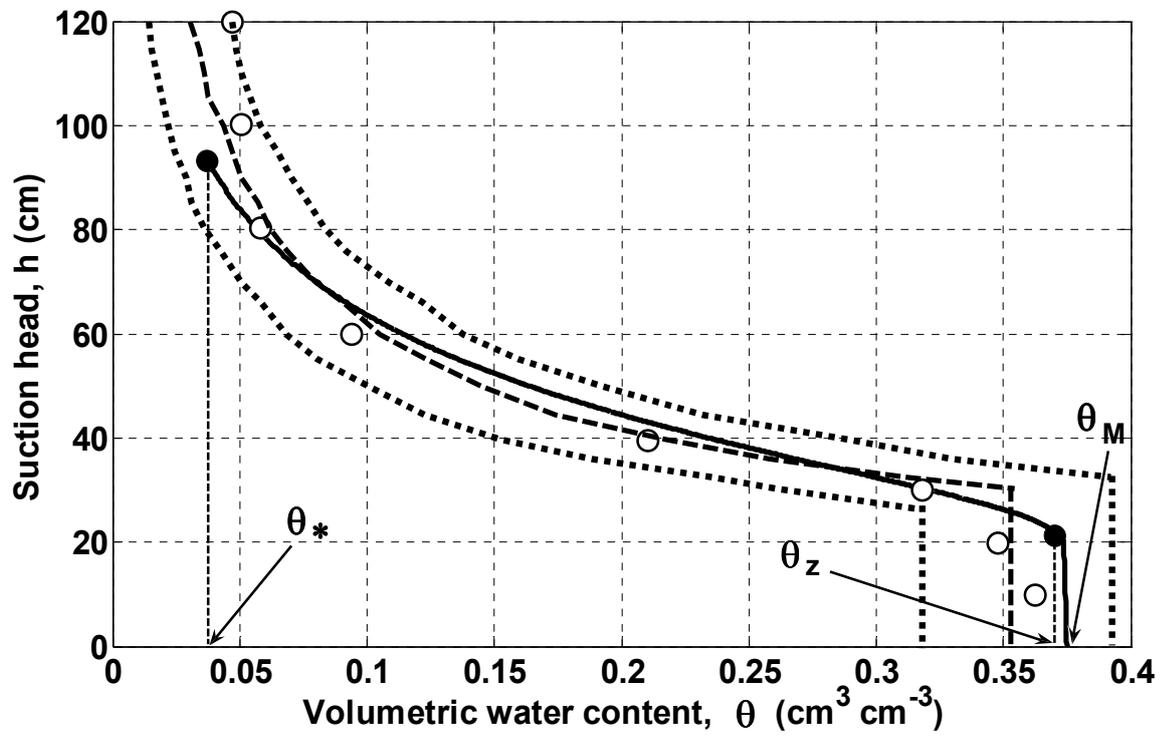

Fig.10

**Table 1. Illustrative estimates of the $\Delta v_w$ and $\Delta v_w/p$ values at typical $X_m$ and $p$ values[†]**

| $X_m$ | $p$ | $\Delta v_w$ | $\Delta v_w/p$ |
|---|---|---|---|
| mm | | | |
| 2 | | $1.8\ 10^{-3}$ | |
| | 0.3 | | $6\ 10^{-3}$ |
| | 0.5 | | $3.6\ 10^{-3}$ |
| 1 | | $3.6\ 10^{-3}$ | |
| | 0.3 | | $1.2\ 10^{-2}$ |
| | 0.5 | | $7.2\ 10^{-3}$ |
| 0.5 | | $7.2\ 10^{-3}$ | |
| | 0.3 | | $2.4\ 10^{-2}$ |
| | 0.5 | | $1.4\ 10^{-2}$ |
| 0.25 | | $1.4\ 10^{-2}$ | |
| | 0.3 | | $4.8\ 10^{-2}$ |
| | 0.5 | | $2.9\ 10^{-2}$ |
| 0.1 | | $3.6\ 10^{-2}$ | |
| | 0.3 | | $1.2\ 10^{-1}$ |
| | 0.5 | | $7.2\ 10^{-2}$ |
| 0.05 | | $7.2\ 10^{-2}$ | |
| | 0.3 | | $2.4\ 10^{-1}$ |
| | 0.5 | | $1.4\ 10^{-1}$ |

[†] $X_m$, maximum grain size; $p$, sand porosity; $\Delta v_w$, relative additional water volume of rigid grain matrix; $\Delta v_w/p$, the same divided by sand porosity.

**Table 2. Input and estimated parameters of the two-factor model of sand water retention**

| Data source | Fig | Input parameters[†] | | | Model estimated parameters[‡] | | | |
|---|---|---|---|---|---|---|---|---|
| | | $p$ | $X_{min}$ | $X_m$ | $\Delta v_w$ | $\theta_*$ | $\theta_z$ | $\theta_M$ |
| | | | μm | μm | | cm$^3$ cm$^{-3}$ | cm$^3$ cm$^{-3}$ | cm$^3$ cm$^{-3}$ |
| Lamara, Derriche [21]. A dune sand | 7 | 0.321 | 45 | 900 | 0.0040 | 0.032 | 0.317 | 0.324 |
| Elrick et al. [22] Preston sand | 8 | 0.393 | 50 | 1200 | 0.0030 | 0.039 | 0.390 | 0.395 |
| Elrick et al. [22] Gormley sand | 9 | 0.357 | 50 | 1700 | 0.0021 | 0.036 | 0.355 | 0.358 |
| Elrick et al. [22] Bolton sand | 10 | 0.373 | 50 | 1200 | 0.0030 | 0.037 | 0.370 | 0.375 |

[†] $p$, sand porosity; $X_{min}$, minimum grain size; $X_m$, maximum grain size.

[‡] $\Delta v_w$, relative additional water volume per unit volume of dry sand (distributed in the saturated sand matrix as water film along grain surfaces); $\theta_*$, lower boundary of the model applicability in terms of volumetric water content; $\theta_z$, volumetric water content at sand saturation; $\theta_M$, volumetric water content at zero suction.